\documentclass[twocolumn,amsmath,amssymb,floatfix]{revtex4}

\usepackage{bm}
\usepackage{amsmath}
\usepackage{epsf}

\newcommand{\beq}{\begin{equation}}

\newcommand{\eeq}{\end{equation}}
\newcommand{\beqa}{\begin{eqnarray}}
\newcommand{\eeqa}{\end{eqnarray}}

\newcommand{\tEE}{\tilde E \tilde E}

\newcommand{\bn}{\hat{\bf n}}

\newcommand{\bl}{{\bf l}} 
\newcommand{\bL}{{\bf L}}

\def\simlt{\lesssim}

\newcommand{\ApJS}{Astrophys. J Supp.}

\newcommand{\ApJ}{Astrophys. J}
\newcommand{\PRL}{Phys. Rev. Lett.}
\newcommand{\PRD}{Phys. Rev. D}
\newcommand{\MNRAS}{Mon. Not. R. Astron. Soc.}

\newcommand{\AsAs}{Astron. Astrophys.}

\newcommand{\amp}{\& }
\newcommand{\etal}{{\it et al. }}
\newcommand{\aut}[2]{{#2.\ #1,}}
\newcommand{\paut}[2]{{#2.\ #1} and}
\newcommand{\laut}[2]{{#2.\ #1,}}
\newcommand{\refs}[6]{#2, {\bf #3},  {#4} (#5).}

\newcommand{\mybib}[2]{\bibitem{#2}}

\begin{document}

\title{Weak Lensing of the CMB: Sampling Errors on B-modes}
\author{Kendrick M. Smith$^{1}$, Wayne Hu$^{2}$ and Manoj Kaplinghat$^{3}$}
\affiliation{
$^{1}$Department of Physics, University of Chicago, Chicago IL 60637\\
$^{2}$Center for Cosmological Physics, Department of Astronomy and Astrophysics,
and Enrico Fermi Institute, University of Chicago, Chicago IL 60637\\
$^{3}$Department of Physics, One Shields Avenue, University of California,
Davis, California 95616}

\begin{abstract}
\baselineskip 11pt
The $B$ modes generated by the lensing of CMB polarization are a primary target
for the upcoming generation of experiments and can potentially constrain quantities
such as the neutrino mass and  dark energy equation of state.   The net sample variance
on the small scale $B$ modes out to $l=2000$
 exceeds Gaussian expectations by a factor of 10
reflecting the variance of the larger scale lenses that generate them.   It manifests itself
as highly correlated band powers  
with correlation coefficients approaching 70\% for wide bands of $\Delta l/l \sim 0.25$.  
It will double the total variance for experiments that achieve a 
sensitivity of approximately 4 $\mu $K-arcmin and a beam of several arcminutes or better 
This non-Gaussianity must be taken into account in the analysis
of experiments that go beyond first detection.
\end{abstract}
\maketitle

\section{Introduction}

As a step on the road toward the ultimate goal of detecting primordial gravitational
waves, upcoming cosmic microwave background (CMB) polarization experiments
will target the distortion to the acoustic polarization induced by
gravitational lensing.   
As with the polarization induced by gravitational waves, the gravitationally
lensed polarization contains a component with handedness,
the so-called  $B$ mode component \cite{ZalSel98}. 
Unlike gravitational waves, gravitational lensing provides a guaranteed signal.  In the
standard cosmological model, the
predicted amplitude of the $B$ modes can only vary at the tens of percent level within current
constraints \cite{Speetal03}.
Moreover these fine variations provide an opportunity to measure the dark side
of the universe, namely the dark energy and neutrino dependent growth of structure, as well as
another handle on the reionization optical depth \cite{Hu01c,KapKnoSon03}.

Although both the intrinsic distribution
and the density perturbations that lens the CMB are expected to be Gaussian, 
the lensed distribution 
is non-Gaussian at second order in the perturbations.    The non-Gaussianity is
therefore relatively small in the temperature distribution \cite{Hu01}.  However because
the $B$ modes are generated by the lensing itself, its non-Gaussianity is a first order
effect but fortunately one that is precisely calculable.
Gravitational lensing therefore also provides a unique
testing ground for experimentally extracting a non-Gaussian signal in
the presence of foregrounds and systematic errors.  
Ultimately, the non-Gaussianity of the lensed polarization also
 provides the key to mapping the dark matter \cite{HuOka01,HirSel03} and
 hence the separation of the
lensing and gravitational wave $B$ mode components \cite{KnoSon02,KesCooKam02}.

For the upcoming generation of experiments, 
the non-Gaussianity will provide an important source
of uncertainty for power spectrum measurements.  Fisher information studies have shown
that the information contained on cosmological parameters in the $B$ mode power spectrum under the
Gaussian approximation unphysically exceeds that contained in the 
 two underlying Gaussian fields \cite{Hu01c}.

In this paper, we study the origin and quantify the impact of the non-Gaussian sample
variance  on
$B$ mode
power spectra measurements.   The basic reason for the large sample variance is 
that the fluctuations that
 lens the CMB are mainly on degree scales.  All of the arcminute scale $B$ modes
 fluctuate jointly with the lens and so precision measurements
 will require many degree scale patches not simply many arcminute scale patches.

We begin in \S \ref{sec:bmodes} by briefly reviewing the generation of $B$ modes
through gravitational lensing.  We calculate the non-Gaussian sample covariance in
\S \ref{sec:covariance} and explore its impact on measurements in \S \ref{sec:impact}.
We conclude in \S \ref{sec:discussion}.
For illustrative purposes we employ throughout a fiducial cosmology that is
consistent with WMAP determinations: an initial scale invariant spectrum of 
curvature fluctuations with amplitude
$\delta_\zeta =5.07\times 10^{-5}$  ($\sigma_8=0.91$, $\tau=0.17$), a baryon density
$\Omega_bh^2 =0.024$ and a matter density
$\Omega_m h^2 =0.14$ in a flat $\Omega_\Lambda=0.73$ cosmology.

\section{B modes}
\label{sec:bmodes}

Weak lensing by the large-scale structure of the universe remaps
the polarization field or equivalently the
dimensionless Stokes parameters $Q(\bn)$ and $U(\bn)$ as 
\cite{BlaSch87,Ber97,ZalSel98}   
\begin{eqnarray}
\, [Q\pm iU](\bn) & = &  [\tilde Q\pm i \tilde U](\bn + \nabla \phi(\bn))\,, 
\label{eqn:qulens}
\end{eqnarray}
where 
$\bn$ is the direction on the sky,
tildes denote
the unlensed field, and the deflection angle $\nabla \phi$ is the gradient of
the  line of
sight projection of the gravitational potential $\Psi({\bf x},D)$,
\begin{equation}
\phi(\bn) = -2 \int d D\, {(D_s-D) \over D\, D_s} \Psi(D \bn,D)\,,
\end{equation}
where $D$ is the comoving distance along the line of sight in  the 
assumed flat cosmology and $D_s$ denotes the distance to the
last-scattering surface.  In the flat sky approximation, the
Stokes parameters can be decomposed into $E$ and $B$ Fourier modes as
\begin{eqnarray}
[Q\pm i U](\bn) &=& 
		- 
		{\int {d^2 l \over (2\pi)^2}}  [E(\bl)\pm i B(\bl)] e^{\pm 2i\varphi_{\bf l}} e^{i \bl \cdot \bn} \,,
\end{eqnarray} 
where $\bl = (l\cos\varphi_l,l\sin\varphi_l)$ and likewise
\begin{eqnarray}
\phi(\bn) &= & 
		\int { d^2 l \over (2\pi)^2}
		\phi(\bl) e^{i \bl \cdot \bn} \,.
\end{eqnarray}

\begin{figure}[t]
\centerline{\epsfxsize=3.5truein\epsffile{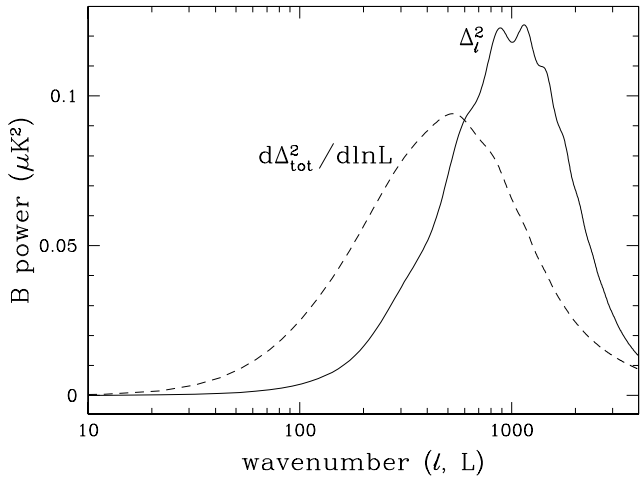}}
\caption{$B$ power spectrum $\Delta_l^2= l^2 C_l^{BB}/2\pi$ and the differential contribution to the total $B$ variance $\Delta_{\rm tot}^2$ from
power in the lensing field $\phi$ with wavenumber $L$ [see Eqn.~(\ref{eqn:deltatot})].  
Even though the $B$ power peaks at $l\sim 10^3$, it gains half its contribution from
power in the lensing field at $L \simlt 450$.  Sampling errors in $B$ therefore reflect the larger
scale variations in $\phi$.}
\label{fig:power}
\end{figure}

Even in the absence of an intrinsic $\tilde B$, lensing
will generate  $B$ as \cite{ZalSel98}
\begin{eqnarray}
B(\bl)      &=&  {\int {d^2 l' \over (2\pi)^2}}
\tilde E(\bl') \phi(\bl-\bl') W(\bl,\bl') \,,
\label{eqn:gradientapprox}
\end{eqnarray}	
where we have expanded Eqn.~(\ref{eqn:qulens}) in the leading order
gradient approximation and the mode coupling weight is
\begin{equation}
W(\bl,\bl') = [\bl' \cdot (\bl - \bl')] \sin 2(\varphi_l - \varphi_{l'})\,.
\end{equation}

Upcoming experiments will mainly measure the power spectrum of the $B$ modes.
In the flat sky approximation, the power spectra of a statistically homogeneous
field $X(\bl)$ is given by
\begin{equation}
\langle  X^*(\bl) X(\bl') \rangle = (2\pi)^2 \delta(\bl - \bl') C_{\bl}\,,
\end{equation}
and statistical isotropy requires $C_{\bl}=C_l$.
It follows that (e.g. \cite{Hu00b})
\begin{equation}
C_l^{BB} = \int {d^2 L \over (2\pi)^2} W^2(\bl,\bl-\bL ) C_{\bl-\bL}^{\tilde E \tilde E} C_{L}^{\phi\phi}\,,
\label{eqn:bbpower}
\end{equation}
given that the intrinsic polarization from $z\sim 1000$
is essentially uncorrelated with the lensing potential from $z \simlt 3$.
The $B$ power spectrum is essentially a convolution of the $\tilde E$ and $\phi$
power spectra.

The total power in the B field arises from power in the $\phi$ field as
\begin{eqnarray}
\label{eqn:deltatot}
\Delta_{\rm tot}^2 &=& \int {d^2 l \over (2\pi)^2} C_l^{BB}  = (0.46\mu {\rm K}/T)^2 \\
& =& 
\int {d^2 L\over (2\pi)^2}
C_{L}^{\phi\phi} \left[ \int {d^2 l \over (2\pi)^2}  W^2(\bl,\bl-\bL )  C_{\bl-\bL}^{\tilde E \tilde E} \right] \nonumber
\end{eqnarray}
where the value in parentheses is the rms in the fiducial cosmology and $T=2.725\times 10^6 \mu$K is
the CMB temperature.  For reference, the rms of the $l \le 2000$ 
low-pass filtered $B$ field is $0.43\mu$K. 
We will typically take this value as the maximum $l$ estimated for illustration purposes.

In Fig.~\ref{fig:power}, we show the $B$ power spectrum in the fiducial cosmology.
Note that the power spectrum peaks at $l\sim 10^3$ reflecting the power in the 
underlying unlensed $\tilde E $ power spectrum.  However this mode coupling
in Eqn.~(\ref{eqn:bbpower}) is achieved through power in the potential $C_L^{\phi\phi}$ 
across a broad range in $L$.  The $B$ field acquires half its power for $L \simlt 450$.   
In other words, degree-scale gravitational lenses
give rise to the $B$ power at the $10'$ scale.  

\begin{figure}[t]
\centerline{\epsfxsize=3.3truein\epsffile{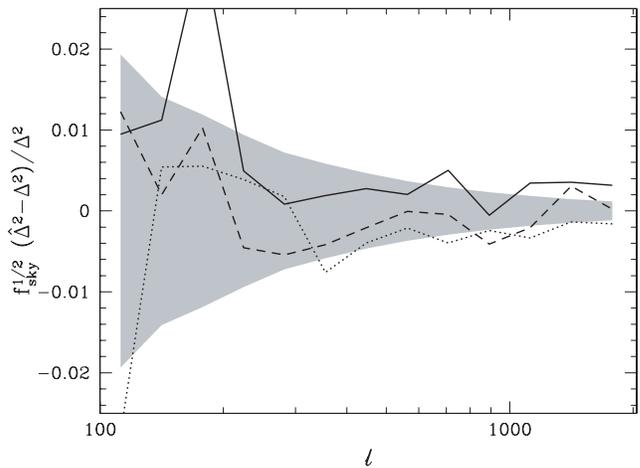}}
\caption{Fractional deviation of band powers from the ensemble average for
three Monte Carlo realizations (lines) compared with the expected rms 
deviation (shaded region) for a Gaussian field of the same power spectrum.  
The realizations show correlated
deviations in the recovered band powers at high $l$.}
\label{fig:realizations}
\end{figure}

This fact is the key to understanding the non-Gaussian covariance.  
In a given degree scale patch of sky there is effectively only a single lens contributing
to the small scale $B$ power.  The amplitude of all of these smaller scale $B$ modes
will covary with the amplitude of the larger scale lens.  Hence reducing the sampling
errors for arcminute scale $B$ modes requires many independent degree 
scale patches of sky.  If the $B$ field were Gaussian,
such a constraint would require
only an
equivalent number of arcminute scale patches of sky.

We illustrate this effect with Monte Carlo realizations of the polarization field and lenses
from the power spectra of the fiducial cosmology.
The Monte Carlo simulations were performed using a square patch of
sky of side length 22.9~degrees, and with fields sampled on a
grid with $1.3$~arcminute spacing.
Periodic boundary conditions were used to eliminate the boundary effects
that would otherwise make 
the decomposition of $Q$ and $U$ into $E$ and $B$ ambiguous.

In Fig.~\ref{fig:realizations}, we show the band powers extracted
from individual runs, using 13 bins logarithmically spaced from
$l=100$ to $l=2000$.
If the covariance were Gaussian, then each estimated band powers
would be uncorrelated from bin to bin, with rms
deviation given by the shaded band.
The non-Gaussian covariance manifests itself here as a
positive correlation from bin to bin, and a higher variance
than one would expect from Gaussian statistics alone.   This behavior is
the Fourier analogue of having
all of the arcminute scale $B$ modes fluctuate jointly in amplitude with
the degree scale lenses.

\section{Sample Covariance}
\label{sec:covariance}


Since the potential fluctuations  are Gaussian on the large
scales that are responsible for lensing 
as are the unlensed $E$-modes
by assumption, the sample covariance can
be accurately quantified analytically.

Consider an ideal, noise-free estimator of the bandpower 
\begin{equation}
\hat \Delta^2_i = {1\over A \alpha_i}
\int_{l \in i} {d^2 l} {l^2 \over 2\pi} B(\bl)B^*(\bl) \,,
\label{eqn:estimator}
\end{equation}
where $A$ is the survey area in steradians and 
\begin{equation}
\alpha_i = \int_{l \in i} {d^2 l }
\end{equation}
is the $l$-space area of the band.  For a flat spectrum 
\begin{equation}
\Delta_l^2 \equiv { l^2 C_l^{BB} \over 2\pi} = {\rm const.}\,,
\end{equation}
it
is an unbiased estimator of the amplitude
\begin{eqnarray}
 \Delta^2_i  &\equiv& \langle \hat \Delta^2_i  \rangle = \Delta_l^2 \,,
\end{eqnarray}
where we have used the relationship $(2\pi)^2 \delta({\bf 0}) = A$ for a finite patch of sky.
In the limit of bandwidths approaching $\Delta l=1$, the relationship holds for any underlying
power spectrum.  The band power weight of $l^2/2\pi$ in Eqn.~(\ref{eqn:estimator})
is appropriate near the peak at $l\sim 10^3$.
For the $l \ll 10^3$ regime where the band power is steeply rising
but the power is nearly flat (see Fig.~\ref{fig:power}), one can drop the weight
factors 
here and still use wide bands.

\begin{figure}[t]
\centerline{\epsfxsize=3.5truein\epsffile{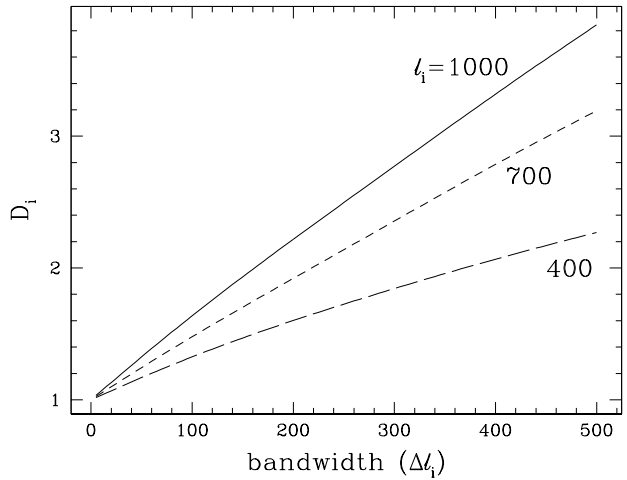}}
\caption{Bandpower variance degradation factor $D_i$ as a function of bandwidth
for various choices of the central $l_i$ of the band.  The degradation increases
with the bandwidth since the Gaussian contribution decreases with the number of
wavemodes.  The trend illustrates the effect of the
non-Gaussian correlation.}
\label{fig:bandwidth}
\end{figure}

Under the assumption of non-overlapping bands in $l$,
the sample covariance of the estimator then follows as
\begin{equation}
S_{ij} \equiv \langle (\hat \Delta^2_i -\Delta^2_i ) (\hat \Delta^2_j-\Delta^2_j ) \rangle
= S_{ij}^G + S_{ij}^N \,.
\label{eqn:cov}
\end{equation}
The first piece is the Gaussian, or more properly the unconnected, contribution
\begin{equation} 
S_{ij}^G=
 \delta_{ij} {2(2\pi)^2\over A \alpha_i^2}
\int d^2 l_i \left( {l_i^2\over 2\pi}
                          C_{l_i}^{BB} \right)^2 \,.
\end{equation}
In the limit of narrow bands and high $l \gg 1$, the Gaussian piece
takes a form that is familiar from Fisher matrix studies \cite{Kno95}
\begin{equation}
S_{ij}^G \approx \delta_{ij}  {2(2\pi)^2 \over A \alpha_i}(\Delta_i^2)^2 
\approx  \delta_{ij}  {2 \over (2l_i+1)\Delta l_i  f_{\rm sky}} (\Delta_i^2)^2 \,,
\label{eqn:gaussiancov}
\end{equation}
where $f_{\rm sky} = A/4\pi$ is the fraction of sky covered.  
The Gaussian errors mainly reflect a mode counting argument.  Since
the fundamental mode  $2\pi/A^{-1/2}$ sets the spacing
of modes in $l$-space, $A \alpha_j/ (2\pi)^2$ is the total number
of modes in the survey area \cite{Teg97} .

The non-Gaussian or connected piece increases the band variances and correlates
them
\begin{eqnarray}
S_{ij}^N &=& {2 \over A\alpha_i\alpha_j}
    \int_{l_i \in i} d^2 l_i  \int_{l_j \in j} d^2l_j \int {d^2 L \over (2\pi)^2}  \nonumber\\
    && \times  {l_i^2l_j^2\over (2\pi)^2} 
    (a_{\bl_i\bl_j}^{\bL} +
    b_{\bl_i\bl_j}^{\bL} +
    c_{\bl_i\bl_j}^{\bL} )
     \nonumber
 \end{eqnarray}
 where
  \begin{eqnarray}
a_{\bl_i\bl_j}^{\bL} &=&
                     W^2(\bl_i,\bl_i-\bL) W^2(\bl_j,\bl_j-\bL)
                     C_{\bl_i-\bL}^{\tEE} C_{\bl_j-\bL}^{\tEE}
                     \big( C_{L}^{\phi\phi} \big)^2 \,,\nonumber\\
b_{\bl_i\bl_j}^{\bL} &=&
                     W^2(\bl_i,\bL) W^2(\bl_j,\bL)
                     \big( C_{L}^{\tEE} \big)^2
                     C_{\bl_i-\bL}^{\phi\phi}
                     C_{\bl_j-\bL}^{\phi\phi} \,, \\
c_{\bl_i\bl_j}^{\bL} &=&
                     W(\bl_i,\bl_i-\bL)W(-\bl_i,\bl_j-\bL)
                     W(\bl_j,\bl_j-\bL)\nonumber \\
                     && \quad 
                     \times W(-\bl_j,\bl_i-\bL)   C_{\bl_i-\bL}^{\tEE} C_{\bl_j-\bL}^{\tEE}
                     C_{\bL}^{\phi\phi}
                     C_{\bl_i+\bl_j-\bL}^{\phi\phi} \nonumber \,.
\end{eqnarray}
Note that the both the Gaussian and non-Gaussian pieces are the same order in the
$C_l^{\tEE}$ and $C_l^{\phi\phi}$ power spectra.

\begin{table*}
\begin{tabular}{c|ccccccccccccc||cc}
$l_i$
   & 112 & 141 & 178 & 224 & 281 & 354
   & 446 & 561 & 707 & 889
   & 1120 & 1409 & 1775 && $D_i$ \\
\hline
112  &  1.00  &  0.03  &  0.04  & 0.05  & 0.05  & 0.06  & 0.06
          &  0.06  &  0.07  &  0.07  &  0.07  &  0.08  &  0.08 && 1.03 (1.04)  \\
141  &  (0.03) & 1.00  &  0.05  &  0.06  &  0.07  &  0.07  &  0.08
          &  0.08  &  0.08  &  0.09  &  0.09  &  0.10  &  0.10 && 1.04 (1.06)  \\
 178   & (0.04) & (0.05) &  1.00  &  0.09  &  0.10  &  0.09  &  0.10 
          &  0.11  &  0.11  &  0.12  &  0.13  &  0.13  &  0.13 && 1.07 (1.10) \\
224   & (0.05) & (0.07) & (0.08) &  1.00  &  0.13  &  0.13  &  0.14
          &  0.15  &  0.15  &  0.17  &  0.17  &  0.18  &  0.18 &&1.13 (1.15)  \\
281   & (0.06) & (0.08) & (0.10) & (0.13) &  1.00  &  0.18  &  0.18 
          &  0.20  &  0.21  &  0.23  &  0.23  &  0.24  &  0.24 &&1.21 (1.23)  \\
354   & (0.05) & (0.08) & (0.09) & (0.13) & (0.19) &  1.00  &  0.22 
          &  0.24  &  0.27  &  0.27  &  0.28  &  0.29  &  0.29 &&1.28 (1.29)  \\
446   & (0.06) & (0.08) & (0.10) & (0.14) & (0.19) & (0.22) &  1.00
          &  0.29  &  0.30  &  0.32  &  0.32  &  0.33  &  0.32 &&1.33 (1.35)  \\
561   & (0.06) & (0.09) & (0.11) & (0.15) & (0.21) & (0.24) & (0.29)
          &  1.00  &  0.37  &  0.40  &  0.40  &  0.40  &  0.39 &&1.55 (1.57)  \\
707   & (0.07) & (0.09) & (0.11) & (0.15) & (0.22) & (0.27) & (0.30)
          & (0.37) &  1.00  &  0.47  &  0.48  &  0.48  &  0.47 &&1.76 (1.78) \\
889  & (0.07) & (0.09) & (0.13) & (0.17) & (0.23) & (0.28) & (0.32)
          & (0.40) & (0.47) &  1.00  &  0.56  &  0.56  &  0.55 &&2.17 (2.17)  \\
1120 & (0.07) & (0.10) & (0.13) & (0.17) & (0.24) & (0.28) & (0.32)
          & (0.40) & (0.47) & (0.55) &  1.00  &  0.62  &  0.61 &&2.56 (2.55)  \\
1409 & (0.08) & (0.10) & (0.13) & (0.17) & (0.24) & (0.29) & (0.32)
          & (0.40) & (0.48) & (0.56) & (0.62) &  1.00  &  0.66 &&2.98 (2.94) \\
1775 & (0.08) & (0.10) & (0.13) & (0.17) & (0.23) & (0.29) & (0.32)
          & (0.40) & (0.46) & (0.54) & (0.60) & (0.66) &  1.00 &&3.19 (3.17) \\
\hline
\end{tabular}
\caption{Correlation matrix
$R_{ij}$ and variance degradation factor $D_i$.  
Parenthetical values are computed from $10^{5}$ Monte-Carlo simulations and 
compare well with the analytic calculation.  
The bands are chosen to be logarithmically spaced and non-overlapping
with $l_{{\rm min}\, i}=0.89 l_i$ and $l_{{\rm max}\, i} = l_{{\rm min}\, i-1}$.}
\label{tab:correlation}
\end{table*}

The non-Gaussianity is conveniently quantified by the variance degradation factor
\begin{equation}
D_i =  {S_{ii} \over S_{ii}^G }\,,
\end{equation}
which gives the diagonals of the covariance matrix,
and the correlation matrix 
\begin{equation}
R_{ij} \equiv {S_{ij} \over \sqrt{ S_{ii} S_{jj} }}\,.
\end{equation}
In Tab.~1 we show these quantities as calculated from the analytic expression
(upper triangle) and $10^5$ Monte-Carlo simulations (lower triangle).
With $10^5$ iterations, the elements of  $R_{ij}$ have converged
to a level ranging from around 0.01 in the lower bins, to around 0.001
in the higher bins (owing to the larger number of modes).
The remaining discrepancies between the analytic and Monte Carlo
results are at the 0.01 level and mainly reflect the use of the gradient
approximation in Eqn.~(\ref{eqn:gradientapprox})
 when deriving the analytic results; the Monte Carlo
simulations resample $Q$ and $U$ as in Eqn.~(\ref{eqn:qulens}).
The close agreement between the two provides confirmation that the
gradient approximation is accurate when computing covariances in the
range $100 \le l \le 2000$.
For these wide bands of $\Delta l/l \sim 0.25$, the non-Gaussian contribution
triples the variance near the peak of the $B$ power and correlates neighboring
bands by 70\%. 

As the bands widen, the non-Gaussian effects superficially appear larger in the
covariance matrix.   In Fig.~\ref{fig:bandwidth}, we show the
degradation factor $D_i$ as a function of the bandwidth.   The nearly linear scaling
at high $D_i$ can be understood as reflecting the linear decrease in the Gaussian errors with 
bandwidth in Eqn.~(\ref{eqn:gaussiancov}) revealing a non-Gaussian floor to the
variance.    For narrow bands, this effect is hidden in the band correlations.
The net effect on the measurements is the same however:
 when combining
 narrow band measurements
 to constrain the parameters underlying the $B$ power spectrum, their non-independence
 leads to a large degradation in the constraints as we show in the next section.

\section{Scientific Impact}
\label{sec:impact}

The sample covariance of band powers provides the ultimate limitations for measuring
the $B$ mode power.  However until experiments map the $B$ field at
 a high signal-to-noise ratio, detector noise will bring the errors closer to Gaussian.

For a given experiment,
the covariance matrix of the band powers will include contributions from detector noise
and the instrumental beam.
The sample covariance matrix of Eqn.~(\ref{eqn:cov}) is replaced with the
full covariance matrix
\begin{equation}
C_{ij} = \langle (\hat \Delta^2_i -\Delta^2_i ) (\hat \Delta^2_j-\Delta^2_j ) \rangle =
C_{ij}^G + S^N_{ij} \,.
\end{equation}
Let us make the usual approximation that that noise is white and Gaussian. Then \cite{Kno95}
\begin{eqnarray}
C_{ij}^G  &=& \delta_{ij} {2(2\pi)^2\over A \alpha_i^2}
\int_{l \in i} d^2 l \left[ {l^2\over 2\pi}
                          (C_{l}^{BB}  + N_{l})\right]^2 \,, \nonumber\\
N_l    & = &
\left( {\Delta_P \over T} \right)^2 e^{l(l+1)\theta_{\rm FWHM}^2/8\ln 2} \,,
\end{eqnarray}
where $\Delta_P^2$ is the polarization noise variance in a steradian and
$\theta$ is the FWHM of the beam in radians.

The impact of the non-Gaussian errors in the power spectrum on 
the errors in a set of cosmological parameters $p_\mu$ can be estimated via the
Fisher matrix
\begin{equation}
F_{\mu\nu} = \sum_{ij}
 {\partial \Delta_i^2 \over \partial p_\mu} 
({\bf C}^{-1})_{ij}
 {\partial \Delta_j^2 \over \partial p_\mu} \,.
 \label{eqn:fisher}
 \end{equation}
 The errors on a given parameter $\sigma(p_{\mu}) \approx ({\bf F}^{-1})_{\mu\mu}^{1/2}$.
Because constraints on the cosmological parameters of interest, e.g. the
neutrino mass and the dark energy properties, will be affected by constraints
from the temperature and $E$ power spectra, we defer a full forecast to
a future work.  Instead we calculate the net degradation in the signal to
noise or equivalently the degradation in the errors of the amplitude of
the $B$ power spectrum given a fixed shape
\begin{equation}
C_l^{BB} = \lambda C_l^{BB} \Big|_{\rm fid} \,.
\end{equation}
For reference, the $B$ power spectrum at the peak $l \sim 10^3$ has a sensitivity to
the dark energy equation of state $w$ of
$d\lambda/dw|_{\Omega_m}\approx-0.6$ and  to the neutrino mass in eV of
$d\lambda/d m_\nu\approx-0.5$.  However these sensitivities should not be used for error
propagation since they carry an $l$-dependence and depend on what is
being fixed by the other spectra.  For example, the sensitivity to $w$ 
at fixed angular diameter distance to last scattering is $d\lambda/dw |_{D_A}\approx -0.2$.
Nonetheless the errors on $\lambda$ serve as a useful quantification of
the overall effect of the non-Gaussianity.

\begin{figure}[t]
\centerline{\epsfxsize=3.5truein\epsffile{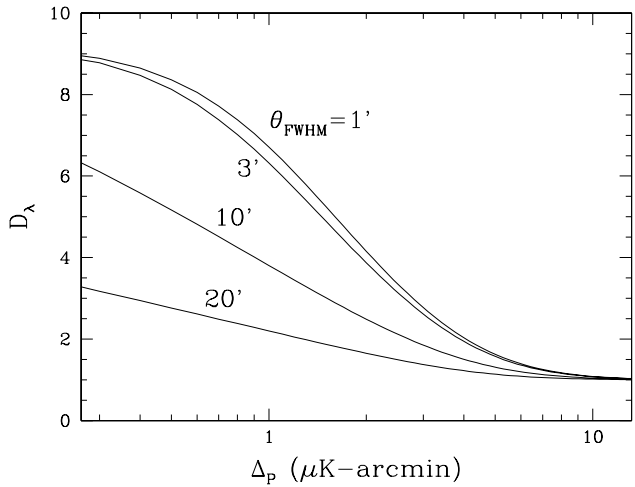}}
\caption{Non-Gaussian degradation on the amplitude $\lambda$
of the $B$ power spectrum
 as a function of detector noise for several choices of the FWHM beam $\theta_{\rm FWHM}$.
For beams that resolve the $B$ power out to the maximum $l_{\rm max}=2000$  
($\theta_{\rm FWHM}\simlt 3'$), the degradation rapidly rises below
$\Delta_P = 4\mu$K-arcmin to the high signal-to-noise, sample variance
 limit  of $\sim 10$.  Increasing $l_{\rm max}$ would further increase
the importance of sample variance.  
}
\label{fig:bottomline}
\end{figure}

From Eqn.~(\ref{eqn:fisher})
the variance in the amplitude becomes
\begin{equation}
\sigma^2(\lambda) = ( \sum_{ij} \Delta_i^2 ({\bf C}^{-1})_{ij} \Delta_j^2 )^{-1} \,.
\label{eqn:generallambda}
\end{equation}
Note that in the Gaussian limit and $\Delta l=1$, the variance becomes
\begin{equation}
\sigma^2(\lambda) \Big|_{G} = 
\left[ \sum_l {2 l +1 \over 2} f_{\rm sky}  \left( {C_l^{BB} \over C_l^{BB} + N_l} \right)^2 \right]^{-1}\,.
\label{eqn:gaussianlambda}
\end{equation}
In Fig.~\ref{fig:bottomline}, we show the degradation factor
\begin{equation}
D_\lambda = {\sigma^2(\lambda) \over \sigma^2(\lambda) |_{G}}
\end{equation}
as a function of the noise $\Delta_P$ in this $\Delta l=1$ limit.
The degradation exceeds a factor of 2 for experiments with $\Delta_P \le 4 \mu$K-arcmin
and a beam $\theta \le 3'$.  At this level, the noise within the beam is comparable
to the rms of the $B$ field [see Eqn.~(\ref{eqn:deltatot})].
 In the limit of zero noise, the degradation is a factor of
$\sim 10$ for bands out to $l=2000$.

  Note that the degradation factor is insensitive to
the choice of banding.  Computing the variance with the wide bands $\Delta l/l  \approx 0.25$
bands of Tab.~1 causes a negligible change in comparison to the accuracy of the
underlying gradient approximation [see Eqn.~(\ref{eqn:gradientapprox})].  
The degradation factor is also
insensitive to the area or $f_{\rm sky}$ for a large contiguous region.  
The degradation can be reduced somewhat
by composing the total area out of many smaller patches separated by many degrees on the
sky.   However this strategy will compromise the separability of $E$ and $B$ modes. 
Optimizing a scan strategy against realistic correlated noise, non-Gaussian sample variance 
and mode leakage will require experiment-specific  Monte-Carlo simulations.

\section{Discussion}
\label{sec:discussion}

The non-Gaussianity of the $B$ modes in the lensed CMB polarization substantially
degrades the amount of information contained in the $B$ mode power spectrum.
It both increases the variance of band powers and makes them strongly covary
across a wide range in $l$ surrounding the peak power.   Ultimately it will increase
the variance of the amplitude of the power spectrum by an order of magnitude.

As experiments move from the upper limit and first detection stage to using
the $B$ mode power spectrum to constrain the properties of dark components
such as the neutrinos and dark energy, this non-Gaussianity will have to be
included in the analysis.  By quantifying the sample covariance, 
we have provided the analytic and numerical tools
that will be the basis for such an analysis.  The advantage of the Monte Carlo
approach is that it can be straightforwardly
applied to any estimator of $B$ power.

In principle, one can include the sample covariance in an effective $\chi^2$ as
is done for power spectrum errors of the temperature field
 (e.g. \cite{Veretal03}).  However since
the computation of the covariance is much more costly than the computation of
the $B$ power spectrum, minimization in a large-dimensional cosmological
parameter space is impractical even with Monte Carlo Markov Chain techniques.   
Since most of the parameters affecting the high redshift universe will be fixed
prior to these measurements from the $T$ and $E$ mode spectra,
as a first order correction one can follow the Fisher matrix approach and
calculate the effect in a fiducial model.  More specifically, given the correlation matrix
 ($R_{ij}$) and the relative variance degradation ($D_{i}$) in a fiducial model,
 one can scale the covariance matrix to the model $B$ mode spectrum 
 ($C_l^{BB}$) as calculated from Boltzmann codes.   This approach would capture
 the main scaling of
 the covariance through the amplitude of the lensing power spectrum.
Implementing such a pipeline though is beyond the scope of this paper.

As experiments move from parameter constraints based on power spectra
to mapping the lensing potential \cite{HuOka01,HirSel03}, 
the non-Gaussianity of the polarization becomes
the signal and not the noise. 
 The extent to which this ultimate goal will be achievable instrumentally
and in the presence of foregrounds  \cite{Bowetal03} awaits the results
of the upcoming generation of experiments.

\smallskip
\noindent{\it Acknowledgments:}  We thank T. Okamoto and B. Winstein for useful discussions.
KMS and WH were supported by NASA NAG5-10840, the DOE and the Packard Foundation.
MK was supported by the NSF and NASA NAG-11098.
It was carried out  in part at  the CfCP under NSF PHY-0114422.

\end{document}